\documentclass[11pt,american]{article}
\usepackage[]{fontenc}
\usepackage{babel}
\usepackage{graphics}
\usepackage{amssymb}
\usepackage[latin1]{inputenc}
\usepackage{times}
\usepackage{mathptm}
\begin{document}

\title{\begin{flushright} \normalsize HD--TVP--99--11, 
accepted for publication in Phys. Rev. Lett.\end{flushright}
\vspace*{0.5cm} Noise induced stability in fluctuating, bistable potentials.}

\author{Andreas Mielke\thanks{
E--mail: mielke@tphys.uni-heidelberg.de
}\\
 Institut f\"{u}r Theoretische Physik, Ruprecht--Karls--Universit\"{a}t, \\
 Philosophenweg 19, D-69120~Heidelberg, F.R.~Germany}

\maketitle
\begin{abstract}
The over-damped motion of a Brownian particle in an asymmetric, bistable, fluctuating
potential shows noise induced stability: For intermediate fluctuation rates
the mean occupancy of minima with an energy above the absolute minimum is enhanced.
The model works as a detector for potential fluctuations being not too fast
and not too slow. This effect occurs due to the different time scales in the
problem. We present a detailed analysis of this effect using the exact solution
of the Fokker-Planck equation for a simple model. Further we show that for not
too fast fluctuations the system can be well described by effective rate equations.
The results of the rate equations agree quantitatively with the exact results.
\\
\\
 PACS-numbers: 05.40.-a, 02.50.Ey, 82.20.Fd, 87.16Ac
\end{abstract}
Models of over-damped Brownian particles in potentials with one or more minima
and barriers serve as paradigms for many relaxation processes in physical, chemical,
and biological systems. The minima represent the stable or metastable states
of the system. Transitions from one state to another are induced by the interaction
with the environment. This interaction is typically described by a thermal white
noise. The dynamics of the system is dominated by characteristic time scales
which are given by the mean first passage times for the escape out of the minima
of the potential. The most simple system in this class of models is the problem
of diffusion over a single potential barrier, pioneered by Kramers \cite{Kramers}. 

In many situations, the potential fluctuates due to some external fluctuations,
chemical reactions, or oscillations. The most prominent model of this kind is
a Brownian particle in a symmetric bistable potential, subject to a harmonic
force. It serves as the standard model for stochastic resonance \cite{SR}.
In many other applications one has to consider stochastic, correlated fluctuations
of the potential. Doering and Gadoua \cite{DoeringGadoua} investigated the
situation of a symmetric, bistable fluctuating potential. They found a local
minimum in the mean first passage time as a function of the barrier fluctuation
rate. This effect has been called resonant activation and has been studied in
detail by various people \cite{Zuercher93}-\cite{Marchi96}. In most of these
papers either a symmetric bistable potential or the escape over a single barrier
has been studied. Escape rates for general potentials and dichotomous as well
as Gaussian fluctuations of the potential have been calculated by Pechukas and
Hänggi \cite{Pechukas94}. Their results support a simple, physical picture
of activated processes with fluctuating barriers: If the potential fluctuates
fast, the rate for transitions over the barrier is determined by the average
barrier. If the potential fluctuations are slow (static limit), the slowest
process determines the rate. In an intermediate regime the rate is given by
the average rate, which is greater than the rate for fast or slow fluctuations.
This picture has already been suggested by Bier and Astumian \cite{Bier93}
on the basis of a simple model with a dichotomously fluctuating linear ramp.

In many applications, one does not have a single barrier or a symmetric, bistable
potential. In a more general situation the potential will have several minima
of different depth. In equilibrium, the system rests most of the time in the
absolute minimum of the potential. But due to potential fluctuations, the position
of the absolute minimum may fluctuate. Typical, biological examples of such
a situation are membrane proteins like a cell surface receptor or an ion channel.
When a ligand binds to the receptor, it changes the potential energy of the
receptor and induces a conformational change of the receptor molecule. If the
transition from one to the other conformation and back is always more or less
the same, a description of this transition by a single coordinate may be sufficient.
Then it is possible to model the conformational changes by the motion of a particle
in a fluctuating potential. 

The effect of a periodic electric field on membrane proteins has been investigated
theoretically \cite{Astumian89a} and experimentally \cite{Petracchi94}. Astumian
and Robertson \cite{Astumian89a} described such a system by a two state model
with periodically modulated rates. The effect of a periodic modulation can be
related to a stochastic, dichotomous modulation \cite{Astumian98b}. This clearly
demonstrates the relevance of our results to such biologically motivated models.
We will come back to this point at the end.

The motion of the over-damped particle in a fluctuating potential can be described
by a Langevin equation 
\begin{equation}
\label{eq:langevin}
\dot{x}=f(x,t)+\xi (t)
\end{equation}
 where \( f=-\frac{\partial V}{\partial x} \). We are using units where the
friction constant and \( k_{B} \) are unity. \( f \) (and \( V \)) depend
on \( t \) since the potential fluctuates. \( \xi (t) \) is a thermal (white)
noise, it satisfies \( \langle \xi \rangle =0 \), \( \langle \xi (t)\xi (t')\rangle =2T\delta (t-t') \).
In this letter, we restrict ourselves to the discussion of potentials with two
minima, separated by a barrier. The position of the minima is \( \pm x_{m} \)
and does not depend on \( t \). The maximum of \( V(x) \) is located at \( x=0 \).
The fluctuation of the potential is mainly a fluctuation of the depth of the
two minima. Let us consider first the most simple, non-trivial version of such
a model. Let us assume that the potential fluctuation is a dichotomous process
and that \( V(x,t) \) takes the two different values \( V_{+}(x) \) and \( V_{-}(x) \).
Further we assume that the absolute minimum of \( V_{+}(x) \) (\( V_{-}(x) \))
is the right (left) minimum. Such a model contains various time scales: Four
mean first passage times for the two minima of \( V_{+}(x) \) and \( V_{-}(x) \),
the intra-well relaxation times for \( V_{+}(x) \) and \( V_{-}(x) \), and
the characteristic time scales for the fluctuation of the potential. The mean
first passage times and the intra-well relaxation times are fixed by the form
of the potential; the fluctuation of the potential is an external parameter
that can be varied. In a biological model for a cell-surface receptor, as mentioned
above, it is determined e.g. by the concentration of the signaling molecule.
Let \( V(x)=\frac{1}{2}(V_{+}(x)+V_{-}(x)) \), \( \Delta V(x)=\frac{1}{2}(V_{+}(x)-V_{-}(x)) \).
Then \( V(x,t)=V(x)+z(t)\Delta V(x) \) where \( z(t) \) is a random process
that takes two values \( \pm 1 \). Its static distribution is \( q_{0}(z)=p_{+}\delta (z-1)+p_{-}\delta (z+1) \).
Let \( \tau  \) be the correlation time of this process, so that \( \langle z(t)z(t')\rangle =\langle z\rangle ^{2}+(1-\langle z\rangle ^{2})\exp (-t/\tau ) \).
Without loss of generality we restrict ourselves to \( p_{-}\le 1/2 \). What
does one expect for such a model? Let us first suppose that the temperature
is such that the typical barrier heights of the system are a few \( T \). If
\( \tau  \) is small compared to the intra-well relaxation times of the potential,
the systems can be described by an effective static potential \( \langle V\rangle (x)=V(x)+\langle z\rangle \Delta V(x) \).
The stationary distribution is \( p_{0}(x)=C\exp (-\left\langle V\right\rangle (x)/T) \).
In the static limit, the stationary distribution is \( p_{0}(x)=p_{+}\exp (-V_{+}(x)/T)+p_{-}\exp (-V_{-}(x)/T) \).
Suppose that \( p_{+} \) is close to unity. Then the average potential is approximately
given by \( V_{+}(x) \) and \( p_{0}(x) \) is approximately the same for small
or large \( \tau  \). In the following we will show that between these two
extreme situations an interesting effect occurs: The mean occupancy of the minimum
at \( -x_{m} \), which is the minimum that has most of the time the higher
energy, may become very large. We will show that this effect is related to resonant
activation.

To calculate \( p_{0}(x) \) or dynamic quantities of the system one has to
solve the Fokker-Planck equation 
\begin{equation}
\label{eq:fp}
\frac{\partial \rho (x,z,t)}{\partial t}=-\frac{\partial }{\partial x}\left( f(x,z)-T\frac{\partial }{\partial x}\right) \rho (x,z,t)+M_{z}\rho (x,z,t).
\end{equation}
 for this model. Here we assumed that the potential fluctuations can be parameterized
by a single stochastic variable \( z(t) \). \( \rho (x,z,t) \) is the joint
probability density for the stochastic variables \( x \) and \( z \), and
\( M_{z} \) is the generator of the stochastic process \( z(t) \). To obtain
the stationary distribution \( p_{0}(x)=\int dz\rho (x,z) \), it is sufficient
to analyze the stationary Fokker-Planck equation. A standard way to solve this
equation is to expand \( \rho (x,z) \) in the right eigenbasis of \( M_{z} \).
If the potential is piecewise linear, one obtains a set of coupled differential
equations with constant coefficients, which can be solved analytically. The
remaining task is to satisfy the continuity conditions for \( \rho (x,z) \),
which is a simple linear algebraic problem. 
\begin{figure}
{\par\centering \resizebox*{0.6\textwidth}{0.3\textheight}{\rotatebox{270}{\includegraphics{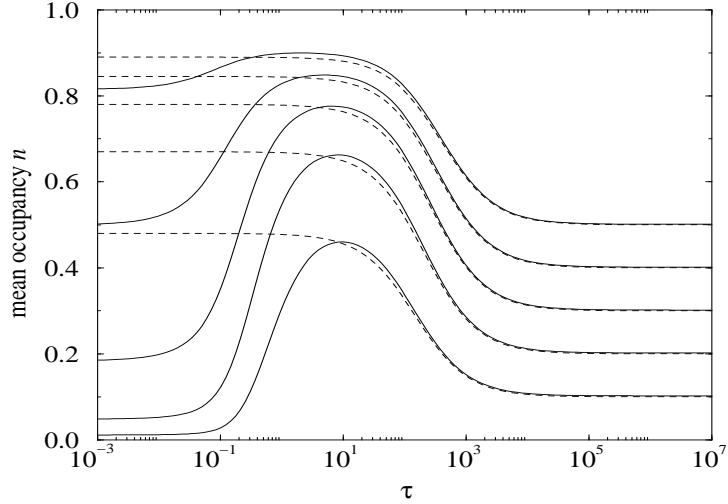}}} \par}

\caption{The mean occupancy in the left minimum of a piecewise linear, dichotomously
fluctuating potential as a function of \protect\( \tau \protect \) for various
\protect\( p_{+}\protect \). The parameters are \protect\( T=0.3\protect \),
\protect\( p_{+}=0.5,\, 0.6,\, \ldots \, 0.9\protect \) , \protect\( f_{+}(x)\protect \)
is piecewise constant and takes the values 10, 1, -1, 2, -2, -10; \protect\( f_{-}(x)\protect \)
takes the values 10, 2, -2, 0.5, -0.5, -10. The values of \protect\( x\protect \)
where \protect\( f_{\pm }(x)\protect \) jumps are -4, -2, 0, 2, 4. The dashed
lines are the results from (\ref{eq:effrate}).}
\end{figure}
For the case of a dichotomous process, the force is \( f(x,z)=f(x)+z\Delta f(x) \)
as discussed above. In Fig. 1 we show results for the probability \( \bar{n}=\int _{-\infty }^{0}p_{0}(x)dx \)
of the particle to sit in the left minimum of the fluctuating potential as a
function of the correlation time \( \tau  \), and for various values of \( p_{+} \).
The choice of the potential is arbitrary, similar results can be obtained for
other potentials as well. The results for \( \bar{n} \) show that the qualitative
discussion for small and large \( \tau  \) given above is valid. But for intermediate
\( \tau  \), \( \bar{n} \) is much larger than expected. The system is able
to detect fluctuations that are not too fast or not too slow. Such fluctuations
enhance the occupancy of the left minimum, although it is most of the time not
the absolute minimum of the potential. We thus observe a noise induced stability
for the state which has most of the time the higher energy, at least when \( p_{+}>1/2 \).
The results in Fig. 1 show that for large values of \( p_{+} \) this effect
is even stronger. 

What is the reason for this effect? How can it be described quantitatively and
how does it depend on the potential? To answer these questions, let us go back
to the general case (\ref{eq:fp}). If \( \tau  \) is large compared to the
intra-well relaxation times for the two minima of \( V(x,z) \), the dynamics
of the system can be described by an effective rate equation for the probability
of the particle to sit in the left minimum, \( n \), or in the right minimum,
\( 1-n \). The rate equation is given by 
\begin{equation}
\label{eq:rate2}
\frac{dn}{dt}=-r_{1}(z)n+r_{2}(z)(1-n)
\end{equation}
where \( r_{1,2}(z) \) is the escape rate for the left or right minimum of
\( V(x,z) \). If \( \Delta V_{1,2}(z)=V(0,z)-V(\mp x_{m},z) \) is the depth
of the potential, \( r_{1,2}(z)\propto \exp (\Delta V_{1,2}(z)/T) \). For (\ref{eq:rate2})
we can again discuss the two limiting cases of large or small \( \tau  \).
For large \( \tau  \), the average occupancy is \( \bar{n}(\tau \rightarrow \infty )=n_{\infty }=\int dz\, q_{0}(z)n(z) \)
where \( n(z)=r_{1}(z)/(r_{1}(z)+r_{2}(z)) \). For small \( \tau  \), the
particle feels average rates and the mean occupancy is \( \bar{n}(\tau =0)=n_{0}=\overline{r_{1}}/(\overline{r_{1}}+\overline{r_{2}}) \)
where \( \overline{r_{i}}=\int dz\, r_{i}(z)q_{0}(z) \). For the results presented
in Fig. 1 the potential has been chosen such that \( n_{0} \) is larger than
\( n_{\infty } \) and also larger than the occupancy determined by the average
potential \( \left\langle V\right\rangle (x) \). This explains qualitatively
the \( \tau  \)-dependence of \( \bar{n} \) in Fig. 1. Let us now compare
the results of the rate equation (\ref{eq:rate2}) with the results of the Fokker-Planck
equation quantitatively. To calculate the stationary probability \( \bar{n} \)
as a function of \( \tau  \) from (\ref{eq:rate2}), we use the Fokker-Planck
equation for the density \( p(n,z,t) \),
\begin{equation}
\label{eq:FokkerPlanck}
\frac{\partial p(n,z,t)}{\partial t}=\frac{\partial }{\partial n}((r_{1}(z)+r_{2}(z))n-r_{2}(z))p(n,z,t)+M_{z}p(n,z,t).
\end{equation}
 The stationary probability \( \bar{n}=\int _{0}^{1}dn\, np_{0}(n) \) can be
obtained from the stationary distribution \( p(n,z) \). It is possible to calculate
the stationary distribution \( p_{0}(n)=\int dz\, p(n,z) \) for a dichotomous
process \cite{AM},
\begin{equation}
p_{0}(n)=C(n-\tilde{n})(n-n_{-})^{\alpha _{-}-1}(n_{+}-n)^{\alpha _{+}-1}
\end{equation}
 where
\begin{equation}
\alpha _{\pm }=\frac{(p_{+}r_{+}+p_{-}r_{-})(n_{\pm }-n_{0})}{\tau r_{+}r_{-}(n_{+}-n_{-})},
\end{equation}
 
\begin{equation}
\tilde{n}=\frac{r_{2+}-r_{2-}}{r_{+}-r_{-}}.
\end{equation}
\( C \) is a normalization constant. We introduced \( r_{\pm }=r_{1\pm }+r_{2\pm } \).
\( r_{i\pm } \) are the two values for the fluctuating rates \( r_{i}(z) \),
and \( n_{\pm }=r_{2\pm }/r_{\pm } \). The rates are given by (see \cite{Bier93},
where \( T \) and \( L \) are set to unity. \( L_{\rm eff} \) occurs instead
of \( L \), since we do not have a linear ramp.)
\begin{equation}
\label{eq:rates}
r_{i\pm }=\frac{\Delta V_{i\pm }^{2}L_{\rm eff}^{2}}{T}(\exp (\Delta V_{i\pm }/T)-\Delta V_{i\pm }/T-1)^{-1}.
\end{equation}
 \( p_{0}(n) \) vanishes outside the interval between \( n_{-} \) and \( n_{+} \)
as it should have been expected. \( \tilde{n} \) does not lie in this interval.
For \( \bar{n} \) one obtains 
\begin{equation}
\label{eq:effrate}
\bar{n}=n_{0}+(n_{\infty }-n_{0})\frac{\tau }{\bar{\tau }+\tau }
\end{equation}
 where \( \bar{\tau }=\frac{p_{+}}{r_{-}}+\frac{p_{-}}{r_{+}} \). This shows
that for a dichotomous process one always has a monotonic behavior of \( \bar{n} \)
as a function of \( \tau  \) and the characteristic time-scale for the transition
from \( n_{0} \) to \( n_{\infty } \) is given by \( \bar{\tau } \). For
more general noise processes it is possible to calculate \( \bar{n} \) as well.
The calculation is much more involved, but the typical behavior of \( \bar{n} \)
is the same as for the dichotomous case \cite{AM}. In Figs. 1 and 2 
\begin{figure}
{\par\centering \resizebox*{0.6\textwidth}{0.3\textheight}{\rotatebox{270}{\includegraphics{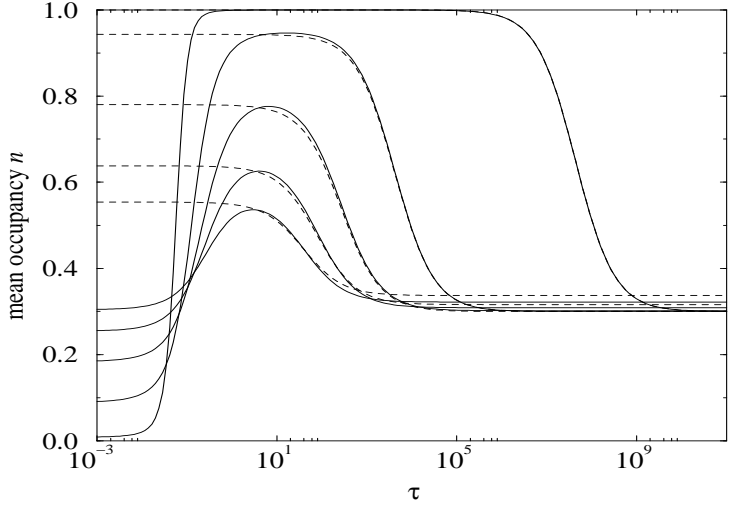}}} \par}

\caption{\protect\( \bar{n}\protect \) as a function of \protect\( \tau \protect \)
for \protect\( p_{+}=0.7\protect \) and \protect\( T=\protect \)0.1, 0.2,
0.3, 0.4, 0.5 (from top to bottom). The potential is the same as in Fig. 1.
The dashed lines are the results from (\ref{eq:effrate}). }
\end{figure}
 we compare results of the effective rate equations with results of the fluctuating
potential. The agreement is indeed excellent for sufficiently large \( \tau  \).
The value of \( \tau  \) where the transition occurs, and the value of the
maximum of \( \bar{n} \) agree well with the exact results. The agreement between
the rate theory and the exact results becomes better for smaller temperatures
(see Fig. 2), which is clear since the rate equations are only valid for low
temperatures. The motion in the average potential, i.e. the behavior of the
system for small \( \tau  \), cannot be described this way, since the rates
in the average potential differ from the averaged rates. The validity of the
two-state model breaks down when \( \tau  \) becomes smaller than the intra-well
relaxation times. Nevertheless, we are able to understand why the occupancy
\( \bar{n} \) in the minimum that has usually the higher energy has the features
shown in Fig. 1. Using the average potential, the average rate, and the average
occupancy we are able to calculate analytically the three values of \( \bar{n} \).
The transition between these values occur at \( \tau  \)-scales given by the
intra-well relaxation time and by \( \bar{\tau } \). This also explains the
results in Fig. 2, which shows how the effect depends on the temperature. For
lower values of the temperature, \( \bar{\tau } \) becomes larger, and the
values of \( n \) change due to the dependence of the rates on \( T \). 

Comparing the above results with calculations for the mean first passage time
shows that the noise induced stability is related to resonant activation. To
calculate the mean first passage time, one has to introduce an absorbing boundary
at the maximum of the potential and has to solve the Fokker-Planck equation
with this boundary. The mean first passage time depends on the initial condition
\( \rho (x,z,0) \), but usually the relaxation within the potential well is
fast compared to the mean first passage time and the dependence on the initial
condition is weak. As initial condition we choose \( \rho (x,z,0)=\delta (x-x_{i})\delta (z-z_{i}) \)
with \( x_{i}=\pm x_{m} \). Let \( x=0 \) be the absorbing boundary. The solution
of the Fokker-Planck equation is denoted by \( \rho (x,z,t|x_{i},z_{i},0) \).
The mean first passage time is then given by \cite{Risken}
\begin{equation}
\label{eq:MFPT}
\tau _{\textrm{MFPT}}=\int dz\int _{-\infty }^{0}dx\rho _{1}(x,z|x_{i})
\end{equation}
 where 
\begin{equation}
\rho _{1}(x,z|x_{i})=-\left\langle \int _{0}^{\infty }t\frac{\partial }{\partial t}\rho (x,z,t|x_{i},z_{i},0)\right\rangle _{z_{i}}.
\end{equation}
 The average is taken with respect to the stationary distribution of \( z_{i} \).
For a piecewise linear potential \( \rho _{1}(x,z|x_{i}) \) can be calculated
using the same methods as for \( \rho (x,z) \) described above. Bier and Astumian
\cite{Bier93} calculated the mean first passage time for a linear ramp, which
is similar to our situation. They showed that for not too small \( \tau  \)
the system is well described by simple rate equations, as in our case as well.
Using the rate equations one obtains for the mean first passage times (\cite{Bier93},
eq. (15))
\begin{equation}
\label{eq:MFPTd}
\tau _{\textrm{MFPT}}(\tau )=\tau _{0i}+(\tau _{\infty i}-\tau _{0i})\frac{\tau }{\tau _{\infty i}+\tau }
\end{equation}
where \( \tau _{0i}=\overline{r_{i}}^{-1} \) and \( \tau _{\infty i}=p_{+}/r_{i-}+p_{-}/r_{i+} \).
The \( \tau  \)-dependence of the mean first passage time is similar to the
\( \tau  \)-dependence of \( \bar{n} \) in (\ref{eq:effrate}). The characteristic
time \( \bar{\tau } \) has the same form as \( \tau _{\infty i} \). 

As already pointed out, it is possible to extend our calculations to various
noise processes. The main qualitative features of the system are the same. Our
results show that an asymmetric, fluctuating, bistable system can detect fluctuations
that are not too slow and not too fast. For such fluctuations the occupancy
of the state that has usually the higher energy is enhanced. The results show
that this effect may be very large, depending on the parameters of the system.
For the largest value of \( p_{+} \) in Fig. 1, the mean occupancy in the left
minimum is very small for slow and fast fluctuations, but reaches a large value
for intermediate values of \( \tau  \). If one lowers the temperature or modifies
the potential it is possible to obtain an even larger effect, as shown in Fig.
2.

To some extend noise induced stability can be compared to noise enhanced stability
first found numerically by Dayan \emph{et al} \cite{Dayan92} and observed experimentally
by Mantegna and Spagnolo \cite{Mantegna96}, but there are several differences.
The effect called noise enhanced stability in \cite{Mantegna96} is observed
in a periodically driven system with a single, metastable minimum. The system
remains in the metastable minimum for some time given by the mean first passage
time for the barrier, and the mean first passage time has a maximum at some
noise intensity. This effect is related to stochastic resonance. In our case
the potential fluctuates stochastically with some correlation time \( \tau  \)
and has two minima. The less stable minimum is the absolute minimum in some
configurations of the potential, but most of the time this minimum is metastable.
Nevertheless it can be highly occupied. 

As metioned above Astumian and Robertson investigated a two-state model with
periodically modulated rates to describe the effect of an oscillating electric
field on membrane proteins. Their results are in qualitative agreement with
our results for the model with dichotomously fluctuating rates. One should expect
that our results for the motion of a particle in a fluctuating potential, described
by a Fokker-Planck equation, are relevant for such biologically motivated models.
This is important, because the description by a Fokker-Planck equation is much
more general. Furthermore, for large frequencies or small correlation times
the systems feels an average potential that cannot be described by fluctuating
rates.


\begin{thebibliography}{10}
\bibitem{AM}A detailed derivation will be given elsewhere.
\end{thebibliography}
\end{document}